\documentclass[amsmath,amssymb,prl,superscriptaddress,twocolumn,showpacs,longbibliography]{revtex4-1}
\usepackage{graphicx}
\usepackage{dcolumn}
\usepackage[colorlinks=true,linkcolor=blue,citecolor=blue,urlcolor=blue]{hyperref}
\usepackage{stmaryrd}
\usepackage{esvect}
\usepackage[usenames,dvipsnames]{xcolor}

\renewcommand{\vec}[1]{\mathbf{#1}}
\newcommand\dd{\text{d}}

\usepackage{siunitx}
\DeclareSIUnit\mub{\mu_\text{B}}
\usepackage{mathtools}

\DeclarePairedDelimiterX\braket[1]{\langle}{\rangle}{#1}

\begin{document}

\title{Nonlocal orbital magnetism of 3\textit{d} adatoms deposited on the Pt(111) surface}

\author{Sascha Brinker}\email{s.brinker@fz-juelich.de}
\author{Manuel dos Santos Dias}\email{m.dos.santos.dias@fz-juelich.de}
\author{Samir Lounis}\email{s.lounis@fz-juelich.de}
\affiliation{Peter Gr\"{u}nberg Institut and Institute for Advanced Simulation, Forschungszentrum J\"{u}lich \& JARA, 52425 J\"{u}lich, Germany}

\date{\today}

\begin{abstract}
The orbital magnetic moment is still surprisingly not well understood, in contrast to the spin part.
Its description in finite systems, such as isolated atoms and molecules, is not problematic, but it was only recently that a rigorous picture was provided for extended systems.
Here we focus on an intermediate class of systems: magnetic adatoms placed on a non-magnetic surface.
We show that the essential quantity is the ground-state charge current density, in the presence of spin-orbit coupling, and set out its first-principles description.
This is illustrated by studying the magnetism of the surface Pt electrons, induced by the presence of Cr, Mn, Fe, Co and Ni adatoms.
A physically appealing partition of the charge current is introduced.
This reveals that there is an important nonlocal contribution to the orbital moments of the Pt atoms, extending three times as far from each magnetic adatom as the induced spin and local orbital moments.
We find that it is as sizable as the latter, and attribute its origin to a spin-orbital susceptibility of the Pt surface, different from the one responsible for the formation of the local orbital moments.
\end{abstract}

\maketitle

Magnetic impurities bridge real and reciprocal space, endowing their non-magnetic host materials with new properties built from the scattering of the itinerant electrons.
The oscillations of the electron density predicted by Friedel~\cite{friedel_metallic_1958} are a classic embodiment of this paradigm, with the giant magnetic moments induced in Pd and Pt attesting to it~\cite{low_distribution_1966,nieuwenhuys_magnetic_1975,oswald_giant_1986,zeller_large-scale_1993,herrmannsdorfer_magnetic_1996,khajetoorians_spin_2013,bouhassoune_rkky-like_2016-2}.
Friedel oscillations lead to long-ranged Ruderman-Kittel-Kasuya-Yosida interactions~\cite{Meier2008,Zhou2010}.
These stabilize helical magnetic chains, which might host Majorana states on superconductors~\cite{Klinovaja2013,Nadj-Perge2013,Pientka2013,Nadj-Perge2014}, and affect the Kondo screening cloud~\cite{Prueser2014}.
Other examples, such as the anomalous Hall effect~\cite{Nagaosa2010,Yu2010b,Chang2013}, Dzyaloshinskii-Moriya interactions~\cite{Fert1980,Menzel2012,Khajetoorians2016}, and large magnetic anisotropy energies~\cite{gambardella_giant_2003,Hirjibehedin2007,khajetoorians_spin_2013,Donati2014a,Rau2014}, highlight the importance of spin-orbit coupling (SOC).
The induced magnetism of the itinerant electrons plays an important role in all of this, but so far little is known about it. 

The interplay between spin and orbital degrees of freedom also underlies one of the most fundamental magnetic properties, the orbital moment~\cite{Skomski2008,Shindou2001,Tatara2003,Hoffmann2015,Dias2016}.
It can be quantified through the Einstein-de Haas effect, see e.g.~\cite{meyer_experimental_1961}, or with x-ray magnetic circular dichroism (XMCD)~\cite{thole_x-ray_1992,carra_x-ray_1993,chen_experimental_1995}.
The classic picture of the orbital moment is based on a superposition of atomic-like swirling charge currents~\cite{Hirst1997}.
For extended systems this picture is incomplete, as explained by the modern theory of orbital magnetization~\cite{xiao_berry_2005,thonhauser_orbital_2005,souza_dichroic_2008,ceresoli_first-principles_2010,thonhauser_theory_2011,hanke_role_2016}, but the interpretation of the non-atomic-like contribution is subtle.
It is then insightful to bring the physics from reciprocal to real space, by considering magnetic impurities on a non-magnetic surface with strong SOC, and the orbital magnetism they induce on the surrounding itinerant electrons.

In this Letter, we formulate the orbital magnetic moment in real space in terms of the ground-state charge current density, partitioned into local and nonlocal contributions.
While the former corresponds to the often-studied atomic-like orbital angular momentum, the latter is unknown for magnetic nanostructures.
Using first-principles, we systematically study transition-metal single adatoms deposited on Pt(111) surface.
We find that the nonlocal orbital moment is as large as the local one, and surprisingly extends much farther into the substrate.
This defines a nonlocal orbital magnetization cloud three times larger than the well-known spin-polarization cloud hosted by the Pt atoms in the vicinity of magnetic atoms~\cite{khajetoorians_spin_2013,bouhassoune_rkky-like_2016-2,hermenau_gateway_2017}.
The moments induced in the substrate can be rationalized as different types of response of the surface to the presence of the magnetic adatom.
In this way, we also prove that the separation into local and nonlocal contributions is meaningful, and that they have distinct physical origins.
The nonlocal contribution to the orbital moment is thus shown to be as important as the local one, with consequences for fundamental studies and possible technological applications.

\begin{figure*}[t]
  \includegraphics[width=\textwidth]{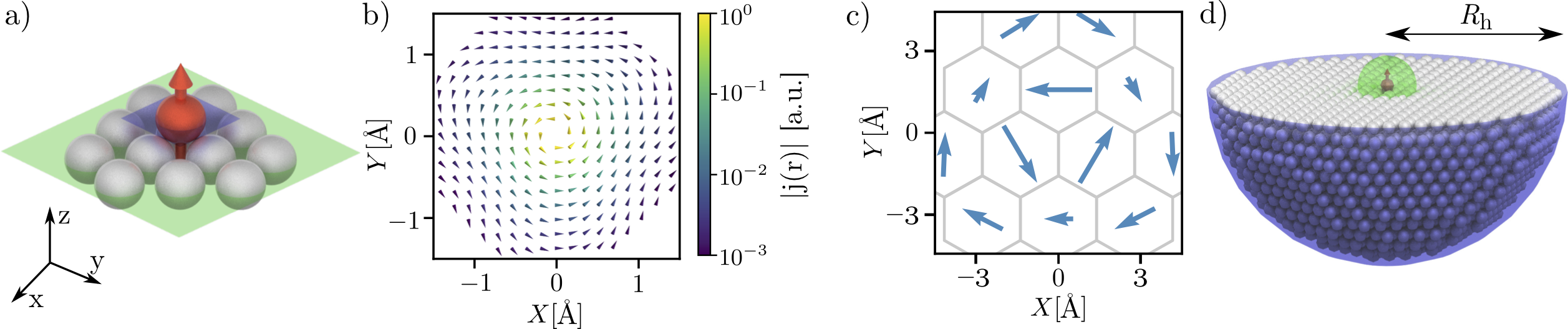}
  \caption{\label{fig:setup_net_currents} 	
  Paramagnetic charge current created by an Fe adatom on the Pt(111) surface.
  (a) Atomic structure. 
  The red sphere represents the adatom, and the grey ones the Pt surface atoms.
  The magnetic moment of the adatom is sketched as a red arrow.
  The adatom is in the fcc hollow position with the vertical distance to the surface reduced to $75\%$ of the bulk interlayer distance.
  The cut planes for panels (b) and (c) are marked in purple and green, respectively.
  (b) Current within the Fe adatom.
  The color scale is logarithmic in atomic units.
  (c) Net charge currents $\vec{j}_i^\text{net}$ in the surface layer of Pt(111).
  These yield the nonlocal contribution to the orbital moment, see Eq.~\eqref{morb_loc_non_loc}.
  (d) Giant cluster consisting of a small central cluster (green sphere) of radius $\SI{2.8}{\angstrom}$, and a large outer one (blue hemisphere) with a radius of $R_\text{h} = \SI{27.2}{\angstrom}$, comprising 2685 Pt atoms.}
\end{figure*}

In classical electrodynamics, the net orbital moment $\vec{m}_\text{o}$ arises from the charge current density $\vec{j}(\vec{r})$~\cite{jackson_classical_1999},
\begin{equation} \label{elementary_def_morb}
  \vec{m}_\text{o} = \frac{1}{2}\!\int\!\dd\vec{r} \; \vec{r} \times \vec{j}(\vec{r}) \quad .
\end{equation}
This formula holds as long as $\vec{j}(\vec{r})$ decays quickly enough towards the boundaries of some enclosing volume.
Microscopically, we can employ the quantum-mechanical ground-state current density, which has three contributions: paramagnetic, diamagnetic, and spin-orbit~\cite{maekawa_spin_2012}.
The paramagnetic contribution is given by
\begin{align}\label{gs_charge_current}
  \vec{j}(\vec{r}) = - \mathrm{i}\,\mu_\text{B} \left[ \Psi^\dagger(\vec{r})\,\big(\vec{\nabla}\Psi(\vec{r})\big)
  - \big(\vec{\nabla}\Psi^\dagger(\vec{r})\big)\,\Psi(\vec{r}) \right] \quad , 
\end{align}
where $\mu_\text{B}$ is the Bohr magneton and $\Psi(\vec{r})$ is the ground-state wave function (written for a single electronic coordinate, for brevity).
The diamagnetic current is absent (no external magnetic fields), and the relativistic correction to the current is found to be small, so it is also neglected.
However, SOC itself is very important, as it
lifts the orbital degeneracy of the surface (needed for a finite ground-state current) via the lifted spin degeneracy due to the presence of the magnetic adatom.

The ground-state paramagnetic current is thus the key to quantify the orbital magnetism induced by the adatom on the surface.
The interpretation is facilitated by partitioning the geometry illustrated in Fig.~\ref{fig:setup_net_currents}(a) into regions centered around each atom, located at $\vec{R}_i$ and with volume $\mathcal{V}_i$.
The orbital moment can then also be split,
\begin{align} \label{morb_loc_non_loc}
  \vec{m}_\text{o} &= \sum_i \frac{1}{2} \left( \vec{R}_i \times \vec{j}_i^\text{net}
  + \int_{\mathcal{V}_i}\hspace{-0.5em}\dd \vec{r}\;\big( \vec{r}- \vec{R}_i \big)\! \times \vec{j}(\vec{r}) \right) \nonumber \\
  &= \sum_i \left( \vec{m}_{\text{o},i}^\text{nl} + \vec{m}_{\text{o},i}^\text{l} \right)
  = \vec{m}_{\text{o}}^\text{nl} + \vec{m}_{\text{o}}^\text{l} \quad .
\end{align}
Eq.~\eqref{morb_loc_non_loc} is independent of the choice of origin (see Supplementary Material~\cite{supp}).
The local contribution $\vec{m}_{\text{o},i}^\text{l}$ captures the swirling of the current around the $i$-th atom, see Fig.~\ref{fig:setup_net_currents}(b), which maps the local orbital angular momentum.
The nonlocal contribution $\vec{m}_{\text{o},i}^\text{nl}$ is due to the net currents $\vec{j}_i^\text{net} = \int_{\mathcal{V}_i}\!\dd \vec{r} \ \vec{j}(\vec{r})$ that flow through the atoms, Fig.~\ref{fig:setup_net_currents}(c).
With Eq.~\eqref{morb_loc_non_loc} we gain access to the spatial dependence of the nonlocal contribution through $\vec{m}_{\text{o},i}^\text{nl}$.

To quantify the ground-state current, we employed a real-space embedding approach based on the Korringa-Kohn-Rostoker Green function method~\cite{papanikolaou_conceptual_2002,Bauer2014,dos_santos_dias_relativistic_2015}, avoiding possible spurious effects due to enforced periodicity.
First the electronic structure of the pristine Pt(111) surface is obtained using a thick slab with open boundary conditions in the stacking direction.
Then clusters of different sizes are self-consistently embedded in the pristine surface, taking into account the relaxation of the adatom towards the surface (see Fig.~\ref{fig:setup_net_currents}(d) for an example).
All quantities can then be systematically converged with respect to the cluster size.
Further computational details can be found in the Supplementary Materials~\cite{supp}.

We first consider the generic features of the paramagnetic charge current, taking an Fe adatom on the Pt(111) surface as exemplary.
The atomic structure is depicted in Fig.~\ref{fig:setup_net_currents}(a).
Fig.~\ref{fig:setup_net_currents}(b) illustrates the current distribution within the adatom.
The current is localized near the nuclear position and swirls mostly in the $xy$-plane, generating a local orbital moment parallel to the spin moment of the adatom.
The adatom also induces a para\-magnetic current in the surface.
Part of it swirls around each Pt atom (not shown), and forms the local orbital moments $\vec{m}_{\text{o},i}^\text{l}$.
The remainder leads to a net current through each Pt atom, Fig.~\ref{fig:setup_net_currents}(c), resulting in the nonlocal contributions $\vec{m}_{\text{o},i}^\text{nl}$.
The direction of the swirl of the net currents is seen to alternate with distance, similar to the  well-known Friedel oscillations of the charge and spin moment~\cite{friedel_metallic_1958,lounis_magnetic_2012,bouhassoune_rkky-like_2016-2}, so $\vec{m}_{\text{o},i}^\text{nl}$ also changes orientation with increasing distance to the adatom.

\begin{figure*}[tb]
  \includegraphics[scale=1]{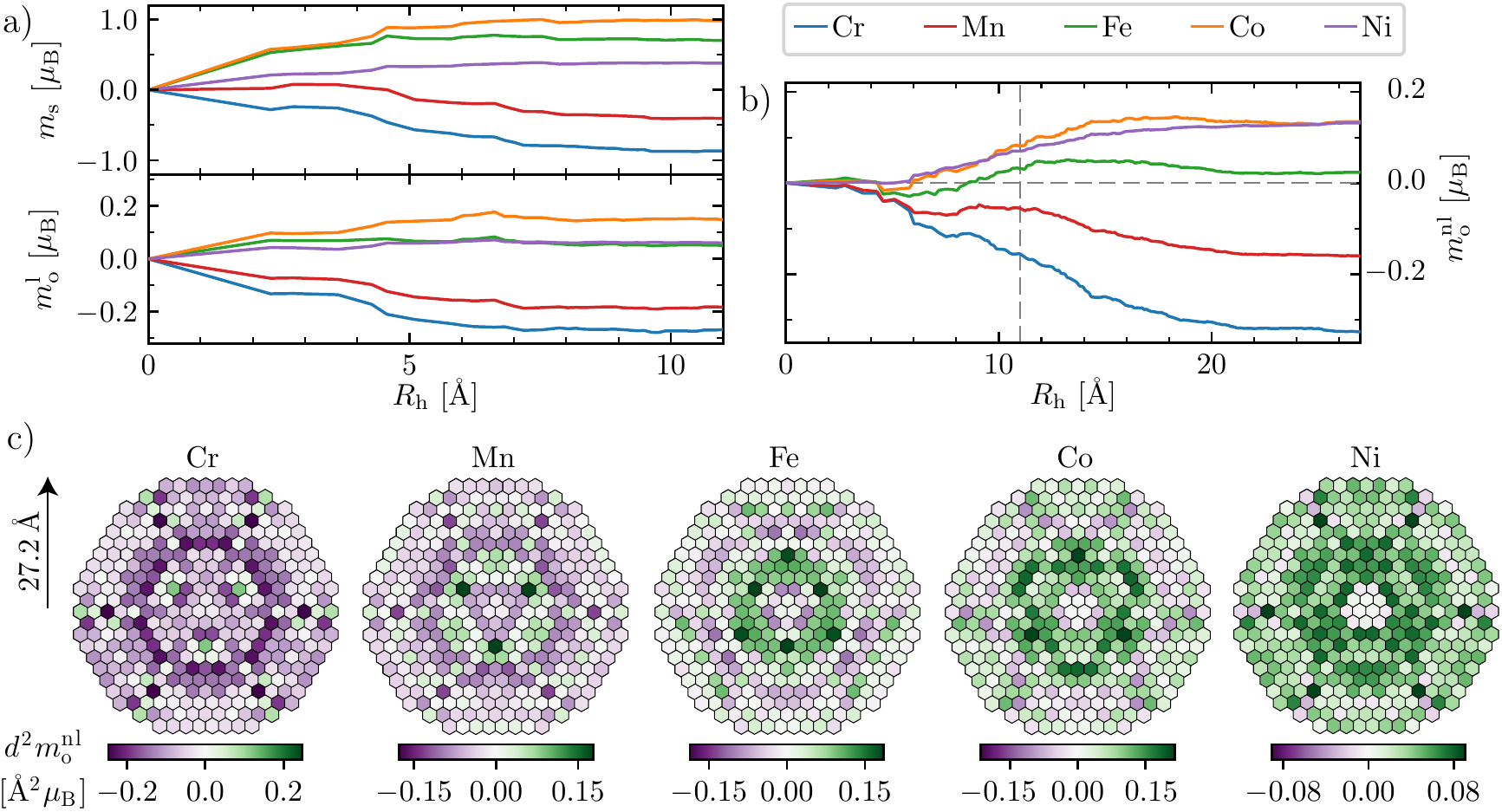}
  \caption{\label{fig:cluster_size_dependence_fcc} 	
  Net induced magnetic moments on the Pt(111) surface due to several adatoms, as a function of the hemispherical cluster radius, $R_\text{h}$.
  (a) Net spin and local orbital moments, $m_{\text{s}}$ and $m_{\text{o}}^{\text{l}}$.
  (b) Net nonlocal orbital moments $m_{\text{o}}^{\text{nl}}$.
  The vertical dashed line at $\SI{11}{\angstrom}$ marks the largest computationally feasible cluster (169 Pt atoms).
  The method described in the main text was utilized to obtain an effective hemispherical cluster with $R_\text{h} = \SI{27.2}{\angstrom}$ (2685 Pt atoms).
  Both methods lead to the same results for $m_{\text{o}}^{\text{nl}}$ (see Supplementary Materials~\cite{supp}).
  (c) Spatial distribution of the nonlocal orbital moments on the Pt surface layer, induced by various adatoms.
  The maps show the component of $\vec{m}_{\text{o},i}^\text{nl}$ normal to the surface.
  The values are scaled by the square of the distance $d$ between a Pt atom and the adatom, showing that they decay faster than $1/d^2$.}
\end{figure*}

Next we investigate the range of the induced magnetic moments, considering various adatoms.
Fixing the spin moment of the adatom to be normal to the surface, the $C_{3v}$ symmetry ensures that the net induced moments are also collinear with it.
Fig.~\ref{fig:cluster_size_dependence_fcc}(a) plots the net spin moments $m_{\text{s}}$ and the net local orbital moments $m_{\text{o}}^{\text{l}}$ against the size of the cluster, showing that they are well-converged beyond $R_{\text{h}} \sim \SI{7.5}{\angstrom}$ (55 Pt atoms).
Surprisingly, the net nonlocal orbital moment $m_{\text{o}}^{\text{nl}}$ has a much longer range than the other two, see
Fig.~\ref{fig:cluster_size_dependence_fcc}(b), going beyond the largest computationally feasible cluster (vertical dashed line).
We tested a physical assumption to overcome the computational limitations: Is the response of a far-away Pt atom to the presence of the magnetic adatom independent of how its nearby Pt atoms respond?
Working with a smaller cluster augmented by a far-away Pt atom and performing calculations for all possible positions of this extra atom, we obtain the response equivalent to that of a giant cluster with 2685 atoms, sketched in Fig.~\ref{fig:setup_net_currents}(d).
We found that the nonlocal orbital moments do follow the previous assumption (see Supplementary Materials~\cite{supp}), validating the results shown in Fig.~\ref{fig:cluster_size_dependence_fcc}(b) beyond the vertical dashed line.
$m_{\text{o}}^{\text{nl}}$ is only converged beyond $R_{\text{h}} \sim \SI{21}{\angstrom}$, showing that it extends about three times as far as the other two contributions to the net induced magnetic moment.

The giant cluster approach can also be used to study the spatial distribution of the nonlocal orbital moments.
These are mapped in Fig.~\ref{fig:cluster_size_dependence_fcc}(c) for the Pt surface atoms, showing Friedel-like oscillations with a fast decay with the distance to a magnetic adatom.
The non-monotonic dependence of $m_{\text{o}}^{\text{nl}}$ on the cluster radius  (see Fig.~\ref{fig:cluster_size_dependence_fcc}(b)) originates from these oscillations.
They are most pronounced for the Fe adatom: the alternating sign of the Pt contributions with increasing distance to the adatom almost cancel each other out when added together.
Surveying the maps for the other adatoms, we see that Ni (Cr) generates mostly positive (negative) contributions to $m_{\text{o}}^{\text{nl}}$, while the oscillations are still present for Co and Mn, although the net contribution is clearly positive for Co and negative for Mn.

\begin{table}[b]
  \begin{ruledtabular}
    \begin{tabular}{cccccc} 
                              &  Cr    &  Mn    &  Fe    &  Co    &  Ni    \\
      \hline 
      $m_\text{s}^\text{ad}$  &  \phantom{-}2.83  &  \phantom{-}3.90  &  3.46  &  2.26  &  0.59  \\
      $m_\text{o}^\text{ad}$  &  \phantom{-}0.05  &  \phantom{-}0.07  &  0.13  &  0.24  &  0.05  \\
      $P^\text{ad}_\text{s}$  & -0.77  & -0.51  &  0.58  &  0.83  &  0.32  \\
      $m_\text{s}$            & -0.87  & -0.41  &  0.70  &  0.98  &  0.38  \\
      $m_\text{o}^\text{l}$   & -0.27  & -0.18  &  0.05  &  0.15  &  0.06  \\
      $m_\text{o}^\text{nl}$  & -0.32  & -0.16  &  0.02  &  0.14  &  0.13  \\
      $\frac{m_\text{o}^\text{nl}}{m_\text{o}^\text{ad}+m_\text{o}^\text{l}}$
                              & 145\%  & 145\%  &  11\%  &  35\%  & 118\%  
    \end{tabular}
  \end{ruledtabular}
  \caption{\label{tab:spin moment_orbital moment}
  Magnetic moments (in $\mu_{\text{B}}$) generated by different adatoms on Pt(111).
  $m_\text{s}^\text{ad}$ and $m_\text{o}^\text{ad}$ are the spin and orbital moments of each adatom, while $P^\text{ad}_\text{s}$ is the relative spin polarization at the Fermi energy of each adatom.
  $m_\text{s}$, $m_\text{o}^\text{l}$ and $m_\text{o}^\text{nl}$ are the spin, local orbital and nonlocal orbital moments of Pt, induced by each adatom.}
\end{table}

The net spin and orbital magnetic moments are collected in Table~\ref{tab:spin moment_orbital moment}.
The spin and orbital moments of the adatoms follow from the filling of their $d$-orbitals.
The spin, local and nonlocal orbital moments induced in Pt are seen to increase when going from Cr to Ni, with an antiparallel alignment for Cr and Mn with respect to the spin moment of the adatom.
The local and nonlocal orbital moments induced in Pt are of similar magnitude.
Furthermore, the nonlocal orbital moment is a substantial part of the net orbital moment for all adatoms, and is even the largest contribution for Cr, Mn and Ni.

\begin{figure}[bt]
  \includegraphics[scale=1]{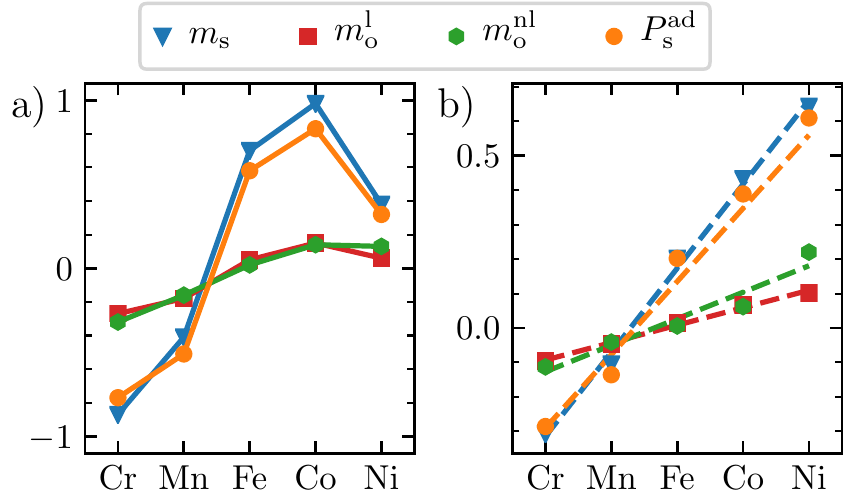}
  \caption{\label{fig:linear_behaviour_2plots} 	
  Relations between the induced surface magnetic moments and the magnetic properties of the adatoms.
  (a) Spin $m_\text{s}$, local $m_{\text{o}}^{\text{l}}$ and nonlocal $m_{\text{o}}^{\text{nl}}$ orbital moments (in $\mu_{\text{B}}$), and relative spin polarization at the Fermi energy $P^\text{ad}_\text{s}$ of each adatom.
  (b) Same as in (a), but with all quantities divided by the corresponding spin moment of each adatom, $m_\text{s}^\text{ad}$.
  The dashed lines are linear fits to the data.}
\end{figure}

Table~\ref{tab:spin moment_orbital moment} also lists the relative spin polarization at the Fermi energy of each adatom, $P^\text{ad}_\text{s} = \frac{\rho_\downarrow(E_\text{F}) - \rho_\uparrow(E_\text{F})}{\rho_\downarrow(E_\text{F}) + \rho_\uparrow(E_\text{F})}$, with $\rho_\downarrow(E_\text{F})$ and $\rho_\uparrow(E_\text{F})$ the minority and majority spin-projected local density of states of each adatom (see Supplementary Materials~\cite{supp}), evaluated at the Fermi energy.
This quantity is closely related with the induced surface magnetic moments, as made apparent in Fig.~\ref{fig:linear_behaviour_2plots}(a).
The remaining ingredient is the spin moment of each adatom, $m^\text{ad}_\text{s}$.
Fig.~\ref{fig:linear_behaviour_2plots}(b) plots the same data but with the values divided by $m^\text{ad}_\text{s}$, resulting in an almost linear correlation between the magnetic moments and $P^\text{ad}_\text{s}$.
The induced magnetic moments are then linear in both $m^\text{ad}_\text{s}$ and $P^\text{ad}_\text{s}$, showing that the surface responds not only to the overall strength of the magnetic perturbation caused by each adatom ($m^\text{ad}_\text{s}$), but also to the spin asymmetry felt by the surface electrons at the Fermi energy ($P^\text{ad}_\text{s}$).

Both $m_{\text{o}}^{\text{l}}$ and $m_{\text{o}}^{\text{nl}}$ arise from the combination of the strong SOC of Pt with the breaking of spin symmetry due to a magnetic adatom.
Pt is well-known to have a large Stoner enhancement of its spin susceptibility, which should be important not just for $m_{\text{s}}$ but also for $m_{\text{o}}^{\text{l}}$ and $m_{\text{o}}^{\text{nl}}$, through its strong SOC.
However, we verified that for $m_{\text{o}}^{\text{nl}}$ the response of a far-away Pt atom is independent of how its nearby Pt atoms respond, while this is false for $m_{\text{o}}^{\text{l}}$ and $m_{\text{s}}$ (see Supplementary Materials~\cite{supp}).
This proves that $m_{\text{o}}^{\text{nl}}$ represents a spin-orbital response of the surface which is distinct from the one leading to $m_{\text{o}}^{\text{l}}$, as could be suspected from their very different spatial range.
A simple explanation is to imagine that each Pt atom responds to the magnetic adatom partly by generating a swirling current centered on it.
This contributes locally to $m_{\text{o}}^{\text{l}}$ but averages out for the local net current, by superposing the contributions generated by the surrounding Pt atoms, leaving $m_{\text{o}}^{\text{nl}}$ unaffected.

We presented a theory of the nonlocal orbital magnetism caused by magnetic nanostructures on a nonmagnetic surface, rooted on knowledge of the induced paramagnetic ground-state current density.
Our detailed study of the magnetism of the Pt(111) surface induced by $3d$ adatoms uncovered several interesting properties of the nonlocal orbital moment: it is long-ranged, displays Friedel-like oscillations, and it arises in a different way than the local orbital moment.
The nonlocal contribution to the orbital moment is as important as the local one, and cannot be neglected.
This is in stark contrast to the case of the elemental bulk ferromagnets, where the nonlocal contribution was found to be small~\cite{Ebert1997,ceresoli_first-principles_2010,hanke_role_2016}.

It remains to be explored whether XMCD measurements employing the sum-rule analysis can detect the full orbital moment of Pt or just part of it, as was argued in a different context in Ref.~\onlinecite{souza_dichroic_2008}.
Scanning probes sensitive to minute spatial variations of the stray fields generated by the surface magnetism might be a future alternative~\cite{Balatsky2012,Rondin2014,Baumann2015}.

The type of surface, the size, shape and dimension of the magnetic nanostructures will matter in defining the magnitude and the decay of the nonlocal orbital moments. Since the latter extends in Pt three times farther than the local orbital and spin moments, nanostructures assumed to be decoupled based on the spatial range of these latter two might actually still be coupled via the former.
This has to be kept in mind when interpreting experimental findings, but could also lead to new long-range interactions mediated by the orbital degrees of freedom, of interest for spinorbitronics.

\begin{acknowledgments}
This work was supported by the European Research Council (ERC) under the European Union's Horizon 2020 research and innovation program (ERC-consolidator grant 681405 -- DYNASORE).
We gratefully acknowledge the computing time granted by the JARA-HPC Vergabegremium and VSR commission on the supercomputer JURECA at Forschungszentrum J\"ulich.
\end{acknowledgments}

\bibliography{Lib_OMM.bib}

\end{document}